\documentstyle[aps,multicol,epsf]{revtex}

\def\u{\uparrow}
\def\d{\downarrow}
\begin{document}
\draft
\title{Spin Driven Jahn-Teller Distortion in a Pyrochlore system}

\author{Yasufumi Yamashita and Kazuo Ueda}
\address{
Institute for Solid State Physics, University of Tokyo, 
Kashiwa-no-ha 5-1-5, Kashiwa-si, Chiba 277-8581, Japan}
\date{Received \today}
\maketitle
\begin{abstract}
The ground-state properties of the spin-1 antiferromagnetic Heisenberg model
on the corner-sharing tetrahedra, pyrochlore lattice,  is investigated.
By breaking up each spin into a pair of 1/2-spins, the problem is reduced
to the equivalent one of the spin-1/2 tetrahedral network in analogy with
the valence bond solid state in one dimension. 
The twofold degeneracy of the spin-singlets of a tetrahedron is lifted
by a Jahn-Teller mechanism, leading to a cubic to tetragonal structural
transition. It is proposed that the present mechanism is responsible for the
phase transition observed in the spin-1 spinel compounds ZnV$_2$O$_4$
and MgV$_2$O$_4$.
\end{abstract}
\pacs{PACS numbers: 75.10.Jm, 75.40.Cx, 75.80.+q}
\begin{multicols}{2}
\narrowtext

Geometrically frustrated spin systems have been a fascinating
subject to study since the Anderson's pioneering work
on classical spins with a disordered ground state\cite{ANDERSON56}.
The pyrochlore and the spinel compounds, both including the
three-dimensional (3D) tetrahedral network sharing the vertexes,
are the typical examples of such systems in nature.
As recent theoretical progress on pyrochlore spin systems,
it may be mentioned that
the spin-1/2 antiferromagnetic (AF) Heisenberg
model is shown to have a spin liquid ground state\cite{LACROIX} 
and that, even for ferromagnetic Ising model, nontrivial effect of frustration 
leads to a newly found spin ice ground state\cite{PYROCHLORE}.
In this letter, we investigate the ground-state properties of the
spin-1 AF Heisenberg model on the pyrochlore lattice.

We follow a general strategy of constructing effective variational wave 
functions, which approximately describe the low energy states
in the same spirit as the resonating valence bond (RVB)\cite{ANDERSON87} 
or the one-dimensional (1D) valence bond solid (VBS) 
approaches\cite{AFFLECK87}.
These simple pictures are of great use to understand 
essential physics of the systems in more intuitive way.
Our simple scenario, where the low energy effective Hilbert space is
assumed to adiabatically continue to the manifold of the 
product wave functions of the twofold degenerate spin-singlets at each
tetrahedron, leads to a spontaneous breakdown of the lattice symmetry.
This new effect of the magneto-elastic interaction in the pyrochlore spin
system gives a consistent picture with the recent experimental results 
on the insulating spin-1 spinel compounds ZnV$_2$O$_4$ 
and MgV$_2$O$_4$\cite{UEDA97,MAMIYA95,MAMIYA97,KONDO,NIZIOL73,PLUM63}.

\begin{figure}[htb] \begin{center}
\leavevmode
\epsfxsize=86mm
\epsffile{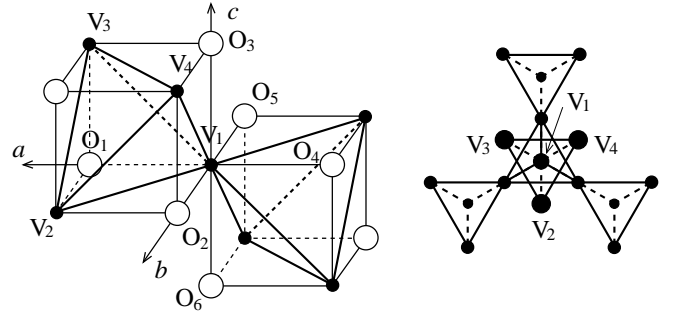}
\caption{The ideal locations of the vanadiums and oxygens in ZnV$_2$O$_4$
and the network of the vanadium ions viewed from the [111] axis.}
\label{fig1}
\end{center} \end{figure}

First, let us introduce the AF Heisenberg model on the pyrochlore lattice,
Fig. \ref{fig1}, in two different but equivalent representations where the
interactions are defined for the sets of bonds and tetrahedra, respectively, 
\begin{eqnarray}
H=J\sum_{<i,j>}\vec{S}_{i}\cdot\vec{S}_{j}
 =J\sum_{k=1}^{N_4}\frac{\left(\vec{S}_k^{tot}\right)^2}{2}+{\rm const.}, 
\quad (J>0),\label{model}
\end{eqnarray}
where $\vec{S}$ is the spin-1 operator and $<i,j>$ denotes
a nearest neighbor pair.
The index $k$, numbered from 1 to $N_4$, 
specifies a tetrahedron and $\vec{S}_k^{tot}$ represents the sum
of the four spins on the $k$-th tetrahedron.
For the present spinel compounds, we expect that the 
relatively small spin value makes the isotropic limit 
a good starting point. On the other hand, it is well known that
there are strong anisotropies for some rare earth titanates\cite{PYROCHLORE}.

To tackle this problem, we develop an analogous method with
that of the VBS state, which captures essential physics of
the Haldane phase in 1D spin-1 systems.
By breaking up the original spin-1 into two spin-1/2 objects
(denoted by $\vec{s}$), we rewrite the Hamiltonian neglecting a constant 
energy shift as follows,
\begin{eqnarray}
H={\cal P}\left\{4J
\sum_{k=1}^{N_4}\frac{(\vec{s}_{k_1}+\vec{s}_{k_2}+\vec{s}_{k_3}+
\vec{s}_{k_4})^2}{2}\right\}
{\cal P},\label{effective}
\end{eqnarray}
where $(\vec{s}_{k_1},\vec{s}_{k_2},\vec{s}_{k_3},\vec{s}_{k_4})$ 
are the spin-1/2 operators forming a tetrahedron. 
${\cal P}$ is the operator to symmetrize the states spanned for all pairs
of the 1/2-spins on every vertex and thus restricts the expanded Hilbert
space to that of the original spin-1 model.
The four times larger coupling constants are required so as to describe 
the original exchange interactions by those within the 1/2-spins
on each tetrahedron.
The tetrahedron-unit representation of the spin-1 pyrochlore system has a
significant difference from the 1D VBS state in the sense that our
fundamental unit is not a bond with two 1/2-spins but a tetrahedron with
four 1/2-spins\cite{KOTOV}. In particular, there are twofold
ground-state degeneracy which makes the situation more interesting
as discussed below.
We believe that this simple transformation is useful
to grasp the essence of the low energy physics of the original problem.

Following the general recipe for the 1D VBS state, we first discuss
properties of the fundamental unit. Since the total spin 
of the tetrahedron is a good quantum number, the two spin-singlets
constitute the degenerate ground states. 
By using the linearly dependent 
three singlets whose total sum equals zero,
$|2\rangle\equiv|\overrightarrow{12}{} \overrightarrow{34}\rangle$,
$|3\rangle\equiv|\overrightarrow{13}{} \overrightarrow{42}\rangle$, and
$|4\rangle\equiv|\overrightarrow{14}{} \overrightarrow{23}\rangle$, where
$|\overrightarrow{ab}{} \overrightarrow{cd}\rangle$ is the product
of the two valence bonds, 
$|\overrightarrow{ab}\rangle=
\left(\u_{a}\d_{b}-\d_{a}\u_{b}\right)/\sqrt{2}$
and similarly defined $|\overrightarrow{cd}\rangle$,
we express the tetrahedron singlets in orthonormal bases, 
\begin{eqnarray}
|R\rangle =\frac{\sqrt{2}}{3}\left(|2\rangle+\omega|3\rangle+\omega^{*}
|4\rangle\right), \;\;{\rm and}\;\;
|L\rangle = |R\rangle ^{*},
\end{eqnarray}
where $\omega=\exp{(2\pi i/3)}$ and $\omega^{*}$ its complex conjugate.
These are the bases of the $E$ representation of the $T_d$ group and 
are the eigen states of the $\pm 2\pi/3$ rotation around 
the four different trigonal axes.

As a next step, we generate the direct products of the local singlets.
By symmetrizing the two 1/2-spins on every vertexes, we obtain the states
\begin{eqnarray}|\psi_{gs}\rangle={\cal P}
\prod_{k=1}^{N_4}\left(r_k|R_k\rangle+l_k|L_k\rangle\right),\label{pregs}
\end{eqnarray}
defined for arbitrary $\{r_k,l_k\}$ with $|r_k|^2+|l_k|^2=1$.
These states are the ground states of the Hamiltonian represented by a sum
of the projection operators onto the total spin three 
(represented by $P^3$) 
subspace for the tetrahedron made of four 1-spins.
Because the four 1/2-spins, out of the 
broken-up eight 1/2-spins, form a singlet state, then the maximum total spin
of the original tetrahedron must be less than three.
To be explicit, 
the projection operator $P^3$ of the $k$-th tetrahedron is given by, 
{\small
\begin{eqnarray}
P^3_k\!=\!
\frac{1}{24}\left(S^{tot}_k\right)^2
\!\!-\!\frac{43}{1440}\left(S^{tot}_k\right)^4
\!\!+\!\frac{7}{1440}\left(S^{tot}_k\right)^6
\!\!-\!\frac{1}{5760}\left(S^{tot}_k\right)^8
\end{eqnarray}}
We assume that the Hamiltonian (\ref{model}) and $\sum_k P^3_k$ belong to the 
same universality class and share the essential properties in the same way
as the relation between the AF Heisenberg model and the VBS model in one 
dimension\cite{KENNEDY90}.
This assumption leads to a consistent picture for the properties of
ZnV$_2$O$_4$ as minutely discussed in the latter part of this letter.
In the above argument, we supposed that the symmetrized states do not vanish.
As a matter of fact, one can show by using the Schwinger boson representation
that the direct products of any linear combination of the form,
eq.(\ref{pregs}), have the same nonzero norm even after the symmetrization.
Accordingly, although the exact dimension of the symmetrized states
is not known, a macroscopic ground-state degeneracy of order
$\sim 2^{N_4}$ is expected for the model given by the projection operator.
It is the key assumption of the  present study that the low energy part of the
Hilbert space of eq.(\ref{model}) adiabatically continues to these symmetrized 
states of the singlets' products.

In real materials, the ground-state entropy must be zero and
it is quite natural to expect that the above-mentioned macroscopic
degeneracy is lifted. In what follows, therefore,
we concentrate on finding
some reasonable mechanism which stabilizes the observed physical state
among a number of nearly degenerate ground states.
As long as the lattice symmetry remains to be cubic, it would be appropriate
to treat the problem as a pure spin system and investigate the effects of the
longer-range interactions as well as the lattice topology as a source of
lifting the degeneracy.
One can argue that this type of lifting produces a kind of chiral
ordering, which will be published elsewhere\cite{chirality}.
When the interaction between the lattice and spin degrees of
freedom is more important, another way of lifting would
take place with spontaneous breaking of the cubic lattice symmetry\cite{TERAO}.
The energy loss from the lattice rigidity and gain from the magneto-elastic
interaction determine the stable structure, a generalized Jahn-Teller effect
of a novel type.  Thus it has some similarity with the
spin-Peierls transition usually discussed in 1D.

\begin{figure}[htb] \begin{center}
\leavevmode
\epsfxsize=86mm
\epsffile{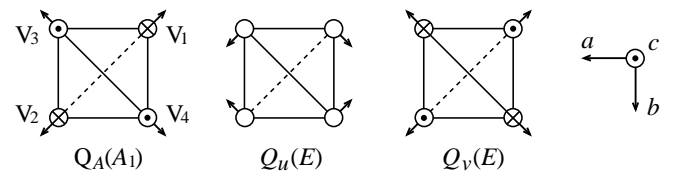}
\caption{The schematic representations of the normal modes 
viewed from the $c$ axis.}\label{mode}
\end{center}\end{figure}

For this purpose we consider a single tetrahedron with 
four vanadium sites (V$_1$-V$_4$) as shown in Fig. \ref{fig1}.
To describe lattice vibrations around the stationary points ($\vec{R}_i^0$),
let us define small deviations 
by $\vec{R}_i=\vec{R}_i^{0}+{\mit \Delta} \vec{x}_i$.
According to the symmetry of the tetrahedron
($T_d$ point group), the normal modes are classified into the 
$A_1$ ($Q_A$: its normal coordinate), $E$ $(Q_u,Q_v)$, and 
$T_2$ $(Q_{\xi},Q_{\eta},Q_{\zeta})$ representations after eliminating
the uniform translation ($T_2$) and the uniform rotation ($T_1$).
The normal coordinates of $A_1$ and $E$ representations are written as,
\begin{eqnarray}
Q_A(A_1)&=&(X+Y+Z)/\sqrt{3},\\
Q_u(E)&=&(X-Y)/\sqrt{2},\\
Q_v(E)&=&(X+Y-2Z)/\sqrt{6},
\end{eqnarray}
where $X,Y$, and $Z$ are the uniform elongation for $x,y$, and $z$
directions, given by 
$(-{\mit \Delta}x_1+{\mit \Delta}x_2+{\mit \Delta}x_3-{\mit \Delta}x_4)/2$,
$(-{\mit \Delta}y_1+{\mit \Delta}y_2-{\mit \Delta}y_3+{\mit \Delta}y_4)/2$, 
and
$(-{\mit \Delta}z_1-{\mit \Delta}z_2+{\mit \Delta}z_3+{\mit \Delta}z_4)/2$,
respectively, see Fig.\ref{mode}.

From simple group theoretical consideration,
the local Hamiltonian, $H_k$, for a single tetrahedron is given,
up to the first order of deviations, by
\begin{eqnarray}
H_k=\sqrt{6}Js_A&+&g_A\frac{\partial J}{\partial Q_A}\biggr{|}_0
Q_As_A
+g_E\frac{\partial J}{\partial Q_E}\biggr{|}_0
\left(Q_us_u+Q_vs_v\right)\nonumber\\
&+&g_T\frac{\partial J}{\partial Q_{T_2}}\biggr{|}_0
\left(Q_{\xi}s_{\xi}+Q_{\eta}s_{\eta}+Q_{\zeta}s_{\zeta}\right),
\label{1st}
\end{eqnarray}
where $s_{\mit \Gamma}$'s are the bases of the irreducible representations
made from the bilinear combinations of the spin operators,
\begin{eqnarray}
\left(\!\!
\begin{array}{c}
s_A \cr s_u \cr s_v \cr s_{\xi} \cr s_{\eta} \cr s_{\zeta} 
\end{array}
\!\!\right)\!=\!\left(\!\!
\begin{array}{cccccc}
\frac{1}{\sqrt{6}}&\frac{1}{\sqrt{6}}&\frac{1}{\sqrt{6}}&\frac{1}{\sqrt{6}}&
\frac{1}{\sqrt{6}}&\frac{1}{\sqrt{6}}\cr
0&\frac{1}{2}&\frac{-1}{2}&\frac{-1}{2}&\frac{1}{2}&0\cr
\frac{2}{2\sqrt{3}}&\frac{-1}{2\sqrt{3}}&\frac{-1}{2\sqrt{3}}&
\frac{-1}{2\sqrt{3}}&\frac{-1}{2\sqrt{3}}&\frac{2}{2\sqrt{3}}\cr
0&0&\frac{1}{\sqrt{2}}&\frac{-1}{\sqrt{2}}&0&0\cr
0&\frac{1}{\sqrt{2}}&0&0&\frac{-1}{\sqrt{2}}&0\cr
\frac{1}{\sqrt{2}}&0&0&0&0&\frac{-1}{\sqrt{2}}
\end{array}
\!\right)\!
\left(\!
\begin{array}{c}
\vec{s}_1\!\cdot\!\vec{s}_2\cr
\vec{s}_1\!\cdot\!\vec{s}_3\cr
\vec{s}_1\!\cdot\!\vec{s}_4\cr
\vec{s}_2\!\cdot\!\vec{s}_3\cr
\vec{s}_2\!\cdot\!\vec{s}_4\cr
\vec{s}_3\!\cdot\!\vec{s}_4
\end{array}
\!\right).\label{ss}
\end{eqnarray}
To illustrate the coupling constants, let us 
assume for simplicity that the exchange coupling depends only on
the distance between the spins. Then $J=J(r)$ and
the derivatives by each normal coordinate
are given by $2\partial J/\partial r$
($A_1$ mode), $\partial J/\partial r$ ($E$), and
$\sqrt{2}\partial J/\partial r$ 
($T_2$), respectively.

In the ground-state spin-singlet subspace,
the matrix elements for the $T_2$ modes vanish and thus may be neglected.
By including the lattice restoring energy, $H_k$ is given by
\begin{eqnarray}
H_k=-\frac{\sqrt{3}g_E}{2}\frac{\partial J}{\partial Q_E}\biggr{|}_0
\!\left(Q_{u}\sigma_x\!+\!Q_{v}\sigma_z\right)\!+\frac{k_E^2}{2}
\left(Q_u^2\!+ Q_v^2\right).
\end{eqnarray}
Here we have neglected the $Q_A$ mode which simply renormalizes the Heisenberg
coupling constant to $J+\left(\partial J/\partial Q_A\right)Q_A/\sqrt{6}$.
We have used the real basis,
$|u\rangle=\left(|L\rangle + |R\rangle\right)/\sqrt{2}$ and
$|v\rangle=\left(|L\rangle - |R\rangle\right)/\sqrt{2}i$ and
the Pauli matrices are defined by
$\sigma_x=|u\rangle\langle v|+|v\rangle\langle u|$ and
$\sigma_z=|u\rangle\langle u|-|v\rangle\langle v|$.
Concerning the symmetrization $\cal{P}$, 
its effect may be taken as a renormalization factor of the order of unity 
($g_A$, $g_E$, and $g_T$ in eq(\ref{1st})).
In other words, the form of the local Hamiltonian is determined by the
symmetry and the important properties such as the equal coupling constants
for the two components of the $E$ mode and the vanishing of the matrix
elements of the $s_{\xi}$, $s_{\eta}$, and $s_{\zeta}$ 
are generic independent of the symmetrization on {\it every} vertexes.

Up to this order, there is no force to determine $\theta$ of
($Q_v,Q_u$)=($\rho \cos{\theta},\rho \sin{\theta}$),
with $\rho=\sqrt{3}g_E\big|\frac{\partial J}{\partial Q_E}\big|/2k_E^2$.
In order to fix this phase $\theta$, it is necessary to take some anisotropies 
into consideration, such as the anharmonicity of the lattice restoration 
energy\cite{OPIK57} or the higher order deviations of the magnetic exchange 
interaction\cite{BALL58}.
After some calculations including up to the third (for the lattice part) and 
fourth order (the spin part) of the lattice deviations, respectively, 
we find both terms to be proportional to $\cos{3\theta}$,
which means that the elongated or compressed lattice structure
along one of the $a$, $b$, or $c$-axis becomes the lowest states
depending on the sign before the $\cos{3\theta}$ term.
We must distort each tetrahedron cooperatively all over the crystal
without contradictions. For example, a uniform local 
tetrahedral distortion causes a uniform lattice metamorphosis and
an alternating lattice distortion develops another type of lattice distortion.
In any way, the many-fold degeneracy is lifted at this structural transition
and a quenched local tetrahedral structure is realized.

Now let us discuss an application of the present scenario
to the spin-1 normal fcc spinel compounds 
ZnV$_2$O$_4$\cite{UEDA97,MAMIYA95,MAMIYA97,KONDO,NIZIOL73} 
or isostructural MgV$_2$O$_4$\cite{MAMIYA95,MAMIYA97,KONDO,PLUM63}, 
where the magnetic V$^{3+}$ ion, with
$(3d)^2$ electronic state forming $S=1$, resides on every lattice point 
of the pyrochlore lattice.
These two compounds show similar properties and we will 
discuss ZnV$_2$O$_4$ mainly.
At $T_{\rm st}=$ 50 K, a cubic to tetragonal structural phase transition
is observed\cite{UEDA97,KONDO,NIZIOL73}, 
where the crystal is uniformly compressed along the $c$ axis
with $c/a\sim$ 0.994 without any magnetic ordering\cite{KONDO,UEDA97} .
This lattice structure is just what we expect
by taking $\theta=0$, which leads to $Q_v>0$ and $Q_u=0$ in our theory.
It is important to consider 
the effect of the 
difference of the $u$ parameter ($u=0.260$\cite{MAMIYA95,NIZIOL73,PLUM63}) 
from its ideal point (1/4), which exists even above $T_{\rm st}$.
The distance between vanadium and oxygen, 
$l_{\rm VO}$, is given by $a(1-4{\mit \Delta}u)/4$, and 
$\cos{\angle {\rm VOV}}=-8{\mit \Delta}{u}$ in the lowest order of
${\mit \Delta}{u}$, where ${\mit \Delta u}= u-1/4$. 
When the V-O-V angle is just the right angle, the superexchange path
through the oxygen vanishes and the spin-spin interaction may be  
ferromagnetic (FM) due to the direct exchange. 
The experimental results imply that $\cos^{-1}(-8{\mit \Delta}{u})$
is sufficient for the AF superexchange interaction to 
overcome the direct FM one. By the distortion due to the
$Q_v$ mode, the cosines of the V-O-V in perpendicular and parallel 
to $c$ axis change to $-8{\mit \Delta}{u}-4{\rho}/\sqrt{6}a$ and
$-8{\mit \Delta}{u}+2{\rho}/\sqrt{6}a$, respectively.
The change of the angles means that the coupling constant
$\frac{\partial J}{\partial Q_E}\big{|}_0$ is positive, see eq. (\ref{ss}).
Therefore for a positive $Q_v$, as the experiments indicate,
there is an enhancement of the AF exchange in the perpendicular 
direction to the $c$ axis and $|u\rangle$ spin state is stabilized 
at each tetrahedron.

The experimental magnetic susceptibility, $\chi$, 
shows a weak temperature dependence
above 100 K with a large Weiss constant, 
$\theta _W=420$ K\cite{MAMIYA95}.
Below $T_{\rm clus}=$ 95 K, $\chi$ shows a splitting depending on
the zero-field cooling (ZFC) and field-cooling (FC), which was interpreted
as development of a short range or cluster ordering\cite{UEDA97}.
But its origin is not clear yet.
Below $T_{\rm clus}$, $\chi$ shows only a gradual increase with decreasing 
temperature\cite{UEDA97,MAMIYA95} and
at $T_{\rm st}$, $\chi$ drops with a sharp cusp structure\cite{MAMIYA97}.
This structure may be consistent with the present scenario.
Since the spin-singlet ground states themselves make no contribution to $\chi$,
the spin part concerned with thermally activated triplet
states and the Van-Vleck type orbital part may be relevant to 
the total susceptibility.
The separation of the doubly degenerate spin-singlet at each tetrahedron
(denoted by ${\mit \Delta}_{S=0}(T)$) tends to suppress the spin part and the 
orbital fluctuation in the vertical plane to [111] axis is also suppressed.
Both terms reduce $\chi$ with a sudden drop accompanied by the structural 
transition, which is qualitatively consistent with the experiments.
With decreasing temperature, the increase of ${\mit \Delta}_{S=0}(T)$ and
the temperature dependent part of $\chi$, still slightly increasing around
$T_{st}$, may balance to result in the almost temperature independent $\chi$
below $T_{st}$\cite{UEDA97,MAMIYA97}.

At $T_N=$ 40 K, a magnetic ordering sets in. The AF structure is observed
by neutron diffraction experiment at 4.2 K\cite{NIZIOL73}, 
where 1D AF chains along [110] and [1$\overline{1}$0] 
directions stack one after the other along the $c$-axis with the 
easy magnetization axis parallel to the $c$ axis.
This magnetic structure is consistent with the change of the AF coupling
constants brought by the Jahn-Teller distortion of the $Q_v$ mode.
To understand the AF long range order from 
the 3D VBS-like $|u\rangle$ spin states symmetrized 
on every vertex realized at $T_{st}$,  
it is necessary to consider the effects of higher triplet states.
 
Finally, let us discuss the electronic entropy, $S_{el}$, of ZnV$_2$O$_4$.
According to the trigonal distortion by the positive ${\mit \Delta} u$,
the threefold degenerate $t_{2g}$ states are expected to split into the 
higher $a_{1g}$ state and lower $e_g$ states. This scheme of the splitting
is expected since 
the wave function of the $a_{1g}$ and $e_g$ states are parallel and
perpendicular to the [111] direction, respectively.
On the other hand, for LiV$_2$O$_4$, the LDA+$U$ calculation\cite{ANISIMOV99} 
suggests that $a_{1g}$ state becomes lower than the
$e_g$ to gain the coulomb energy of the singly occupied electron level
of the $(3d)^{1.5}$ configuration,
but this may not be the case for the $(3d)^2$ state.
Suppose that the orbital degree of freedom is almost
quenched at higher temperatures, then the main contribution to the electronic
entropy around the structural transition is the spin degrees of
freedom in the spin singlet sector, which is estimated to
be $R\ln{2}=5.76$ J/mol$\cdot$K and the lattice part additionally.
This is consistent with the experimental value of
$S_{el}\sim$ 7 J/mol$\cdot$K just above the $T_{\rm st}$\cite{KONDO}.

Another possible scenario for the structural transition is, of
course, the usual Jahn-Teller distortion at $T_{\rm st}$. For this
scenario, it is necessary to assume that the $a_{1g}$ state is lower
than the $e_g$ states. However, the entropy associated with the orbital
degrees of freedom is $2R\ln{2}=11.52$ J/mol$\cdot$K,
which is much larger than the entropy observed experimentally.
Note here that the spin degrees of freedom are not included yet.
It is also worth to mention that for the present scenario of the spin-driven
lattice distortion, the energy scale should be a fraction of the
exchange interaction $J\sim 10^2$ K.
The smallness of the $T_{\rm st}$ and structural metamorphosis
at $T_{\rm st}$ are compatible with this small energy scale.
On the other hand, for the conventional Jahn-Teller effect concerned with
the single electron states, the energy scale is typically a fraction of
an electron volt ($\sim 10^4$ K). 

In conclusion we have studied the frustrating quantum spin-1 system on
the pyrochlore lattice as a theoretical model of the insulating spin-1 spinel
compound. 
In this problem, it is essential to lift the many-fold
degenerate spin-singlet manifold to discuss the low energy physics.
As such a mechanism, a magneto-elastic interaction is considered
in this letter and we find the spontaneous breakdown of the lattice symmetry,
which is consistent with the structural phase transition observed in 
ZnV$_2$O$_4$ and MgV$_2$O$_4$.
Following this scenario, we can understand qualitatively the low temperature
behaviors of the magnetic susceptibility and the electronic entropy,
which seems to be difficult to be explained by the usual Jahn-Teller effect.
It is natural to expect that spin-driven structural transition discussed
in the present paper may be generalized to integer-spin pyrochlore systems
with isotropic interactions.
\vspace*{-.3cm}
\acknowledgments
\vspace*{-.3cm}
We are grateful to S. Kondo and M. Sigrist for many valuable discussions.
Thanks are also due to N. Kawakami and C. Itoi for helpful comments.

\end{multicols}
\end{document}